      [1999/12/01 v1.4c Il Nuovo Cimento]
\documentclass{cimento}

\newcommand{\ltsimeq}{\raisebox{-0.6ex}{$\,\stackrel
        {\raisebox{-.2ex}{$\textstyle <$}}{\sim}\,$}}
\usepackage{graphicx}  % got figures? uncomment this
\title{From yeV to TeV: Search for the Neutron Electric Dipole Moment}
\author{
D.~H.~Beck\from{ins:uiuc},
D.~Budker\from{ins:berkeley}\from{ins:LBNL}, and B.~K.~ Park\from{ins:berkeley}\\
\ for the nEDM Collaboration\from{ins:nEDM}
}
\instlist{
\inst{ins:uiuc} Department of Physics, University of Illinois at Urbana-Champaign, Urbana, IL, USA
\inst{ins:berkeley} Department of Physics, University of California Berkeley, Berkeley, CA 94720-7300, USA
\inst{ins:LBNL} Nuclear Science Division, E.~O.~Lawrence National Laboratory, Berkeley, CA 94720, USA
\inst{ins:nEDM} {\tt http://www.phy.ornl.gov/nedm/}
}
\PACSes{\PACSit{14.20.Dh}{Neutron properties}}
\begin{document}

\maketitle

\begin{abstract}
The existence of electric dipole moments (EDM) for fundamental particles signals time-reversal symmetry (T) violation accompanied by violation of parity (P); only upper limits have been established to date.  Time-reversal violation in turn implies CP violation under the assumption that CPT is a good symmetry.  The neutron is an attractive system for an
EDM search, both because it is neutral and because a neutron EDM would be relatively easier to interpret than the comparable quantity for a nucleus or even an atom.  We introduce briefly the key experimental requirements for such search and describe some aspects of the neutron EDM experiment planned for the Spallation Neutron Source at the U.S. Oak Ridge National Laboratory.
\end{abstract}

\section{Introduction}

Searches for electric dipole moments (EDM) of fundamental particles have a history that dates back to the 1950's.  In this era, the violation of parity was first postulated and then observed in weak-interaction experiments of Wu, et al.~\cite{wu}, marking the beginning of the demise of discrete symmetries in general (so far, with the exception of CPT invariance). The first EDM search for the neutron was done early in the decade by Smith, Purcell and Ramsey as search for parity violation in a scattering experiment at Oak Ridge.  Although the measurements were completed by 1951, the paper was not published until 1957~\cite{smith}, after the publication of the Lee and Yang paper~\cite{lee} on the weak interaction, because the upper limit determined, $\left | d_{n} \right | \le 3.9 \times 10^{-20}$ e$\cdot$cm (all 90\% C.L.), was not thought to be interesting at the time.

In fact, upper limits on EDM have continued to limit theory in important ways over the intervening 60 years.  The CP violation in the standard model generates very small EDM, typically coming in only at the level of a third order loop~\cite{khriplovich}. Therefore, in the search for new physics, EDM are attractive because the standard model `background' is several orders of magnitude below currently observed limits.  There are two prominent threads that argue for continued searches. First, in 1967 Sakharov argued~\cite{sakharov} that any process that could generate the matter-antimatter asymmetry observed in the universe must involve CP violation; the aforementioned CP violation in the standard model is much too small to account for the observed asymmetry~\cite{dar}.  Second, in the currently popular models of physics beyond the standard model, e.g., SUSY, there is much more opportunity for the necessary complex couplings (analogous to those in the CKM matrix) in the SUSY breaking mechanisms of our low-energy world.  The current set of EDM upper limits has significantly restricted the space of possible SUSY models~\cite{musolf}.  With these experiments, we are therefore exploring aspects of physics beyond the standard model that pertain to the TeV-scale energies, and in ways that complement the direct searches at the LHC.

\section{EDM Experiments}

Most EDM experiments look for precession of a fundamental particle's angular momentum in an external electric field.  The interaction energy of the electric dipole in the electric field combines with that of the T-allowed magnetic moment in an external magnetic field in a way that either increases or decreases the normal precession frequency.  These effects are very small; the interaction energy at the sensitivity limit of our proposed Oak Ridge experiment is about $10^{-23}$~eV or 10 yocto-electron volts (yeV); about ten orders of magnitude smaller than the corresponding magnetic energy.  We next survey a few of the key experiments and techniques in the field.

A landmark EDM experiment was carried out at Berkeley by Eugene Commins' group.  Taking advantage of the relativistic enhancement of atomic EDM in heavy atoms~\cite{sandars}, Commins used an atomic beam of polarized (paramagnetic) thallium to measure an EDM which is mostly sensitive to an EDM of the electron.  It is essentially a variant of the Ramsey separated oscillatory field method~\cite{ramsey} wherein the magnetic moments of optically polarized atoms are flipped by a first $\pi/2$ pulse, precess freely in a region where the magnetic and electric fields are either parallel or anti-parallel, and then are analyzed by applying a second, synchronous $\pi/2$ pulse. The resulting polarization is then probed by the same laser used to produce the initial polarization.  The Commins group measured an upper limit $\left | d_e \right | \le 1.5\times 10^{-27}$ e$\cdot$cm~\cite{commins}.
%It was recently superceded by
Essentially the same limit of $\left | d_e \right | \le 1.05\times 10^{-27}$ e$\cdot$cm \cite{hinds} has been obtained recently by the Hinds group at Imperial College in an experiment otherwise similar, but taking the advantage of the large {\it internal} electric field in the polar YbF molecule.

The record for the smallest EDM upper limit is held by the Hg experiment in Seattle.  In this diamagnetic atom, the atomic EDM is most sensitive to the intrinsic nuclear EDM, and there is enhancement from the finite size of the nucleus.  The experiment uses a pair of cells with the same nominal magnetic field, but opposite electric fields, and a second pair of cells with no electric field acting as a co-magnetometer. The atoms are first optically pumped with a circularly polarized laser beam propagating perpendicular to the magnetic field and modulated at the Larmor frequency.  After the atoms are polarized, the light polarization is switched to linear, and a linear polarizer analyzes optical rotation in the transmitted light induced by the precessing atoms.  The limit on the $^{199}$Hg EDM is $\left | d_{Hg} \right | \le 2.6 \times 10^{-29}$ e$\cdot$cm~\cite{heckel}.

The current limit on the neutron EDM is held by the ILL experiment.  It utilizes ultra-cold neutrons (UCN) with energies $\ltsimeq 110$ neV, which can be trapped by the Fermi potentials of certain materials like SiO$_2$ and diamond-like carbon (the UCN-storage possibility was first envisioned by Zel'dovich in 1959~\cite{zel'dovich}, see also Ref.~\cite{golub_book}). The UCN are produced by slowing cold neutrons via scattering from a receding turbine blade.  After the neutrons are polarized by transmission through a magnetized foil, they precess in a cell with parallel or anti-parallel $\vec E$ and $\vec B$ fields and are analyzed in a Ramsey separated oscillatory field experiment as described above.  The limit set in this experiment is $\left | d_{n}  \right | \le 2.9 \times 10^{-26}$ e$\cdot$cm~\cite{harris}.

New n-EDM experiments are currently being developed for neutron facilities at ILL, PSI, Munich, TRIUMF, and the Oak Ridge Spallation Neutron Source.  All primarily seek to increase the UCN density significantly above the $1$ cm$^{-3}$ in the first ILL experiment.  Additional co-magnetometry is also considered to be an important improvement to help circumvent the geometric-phase systematic uncertainty~\cite{pendlebury_Berry,pendlebury} that ultimately limited the ILL experiment described above.

\section{Neutron EDM experiment at the SNS}

\begin{figure}
\includegraphics[width=0.9\textwidth]{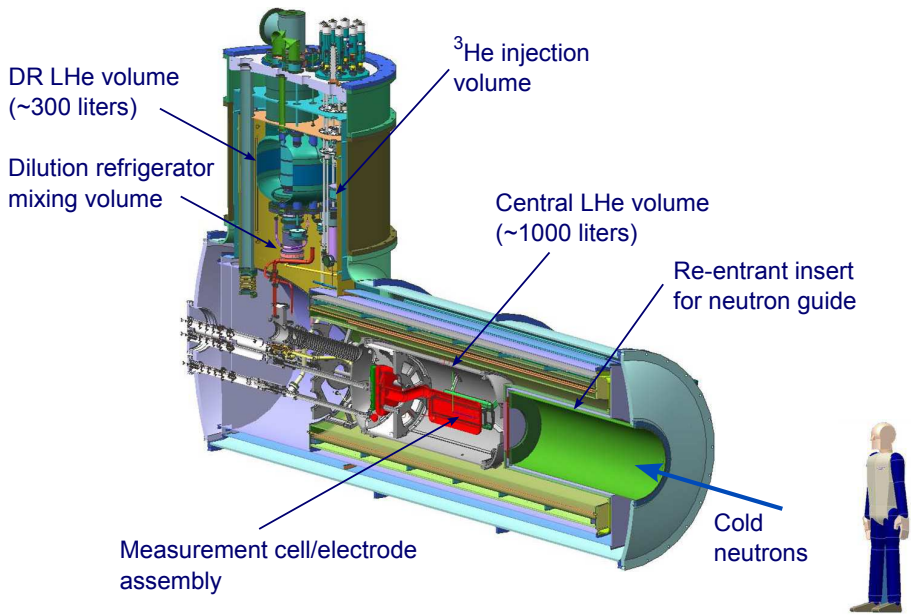}
\caption{The figure shows the nEDM apparatus with details for the liquid helium (LHe) volumes. The measurement cell is located in the central LHe volume; during the measurement cycle, LHe is purified and injected with polarized $^3$He in the dilution refrigerator (DR) LHe volume. \label{fig:apparatus}}
\end{figure}

\subsection{General description}
The sensitivity limit of all measurements of electric dipole moments is, up to a numerical factor
\begin{eqnletter}
\delta d \sim \frac{1}{E} \frac{1}{\sqrt{N \tau T}},
\end{eqnletter}
where $E$ is the electric field, $\tau$ is the coherence time and $T$ the overall measurement time.  The obvious factors to attack are therefore $E$, $N$ and $\tau$.

In order to increase $N$, this experiment (see Fig.~\ref{fig:apparatus}) uses the downscattering of neutrons by resonant creation of the phonon Landau-Feynman excitations in superfluid $^4$He at $T=0.3-0.5$~K.  Cold neutrons with wavelengths of 8.9 \AA, produced either by a graphite monochromator or a chopper, enter the experiment from a standard polarizing guide.  The wavelength corresponds to the crossing of the dispersion curves of the neutrons and superfluid elementary excitations, enabling the neutrons to efficiently give up their energy and momentum to the phonons/rotons~\cite{golub}, which are eventually absorbed by the cold walls of the container.  In fact, thus scattered neutrons are at a temperature much below that of the bath (the upscattering rate is small) and have an energy below that of the Fermi potential associated with the polystyrene coated acrylic container for the superfluid.  In this manner, neutron densities of order 100 cm$^{-3}$ can be produced.

However, even this large number of polarized neutrons, trapped in two three-liter cells of superfluid helium (with opposing electric fields), is neither sufficient for direct detection using SQUID or atomic magnetometers, nor is it amenable to the Ramsey technique. Thus, a second major point of departure for this experiment is the introduction of polarized $^3$He (with a magnetic moment about 10\% larger than that of the neutron) into the superfluid to act as the `detector'~\cite{lamoreaux}.  In this case, we produce polarized $^3$He atoms using an atomic beam source (ABS).  The neutron-capture reaction on $^3$He
\begin{eqnletter}
\hbox{n + $^3$He} \rightarrow \hbox{p + t + 764 keV}
\end{eqnletter}
is highly spin dependent, with a cross section of megabarns when the spins are anti-parallel and effectively zero with spins parallel.  Therefore, by applying a single $\pi/2$ pulse, flipping both the neutron and $^3$He spins, the difference in their precession rates can be measured by observing the scintillation light produced in the liquid helium by the capture products.  Because of the large cross section, the optimal density of $^3$He is low, about $10^{12}$ cm$^{-3}$ corresponding to a $^3$He/$^4$He fraction of about $5\times10^{-11}$, or about four orders of magnitude below the natural concentration.  Even at this low density, however, we can use the overall precession rate of the polarized $^3$He, measured by SQUID, to determine the magnetic field.  Aside from a small gravitational offset (because of the lower neutron temperature and larger $^3$He mass), the neutrons and $^3$He atoms sample exactly the same space inside the superfluid-filled cells, thus an effective co-magnetometer is built-in. To maximize the effectiveness of the co-magnetometer, a spin-dressing technique has been developed and experimentally validated \cite{esler,chu}, in which the gyromagnetic ratios of the neutrons and $^3$He are rendered effectively the same by applying off-resonant radio-frequency fields.

Maximum sensitivity results when measuring for approximately the neutron lifetime, with the capture lifetime tuned (by adjusting the $^3$He density) to about the same value.  Because the $^3$He will eventually be depolarized by wall collisions and field gradients, we must have a technique for removing it from the system and supplying a new charge of highly polarized $^3$He.  For this purpose, we will again use the phonons in the superfluid, this time produced by a heater.  These phonons scatter the $^3$He toward the cold end of the region (see Refs.~\cite{hayden,baym} and references therein) where the heater has produced a temperature gradient.  After dumping the volume containing the concentrated, somewhat depolarized $^3$He, refilling it with pure $^4$He superfluid, a new set of highly polarized $^3$He atoms is injected into the experiment from the ABS.

With a design value for the electric field of 50 kV/cm and a 300-day live-time, this experiment is expected to reach the level of about $8 \times 10^{-28}$ e$\cdot$cm.  The largest systematic uncertainties are expected to be from the geometric-phase effect ($\sim 2 \times 10^{-28}$ e$\cdot$cm, limited by
the uniformity of the $B_0$ holding field), with contributions from an effective magnetic field coming from the neutron scattering from polarized $^3$He, as well as from leakage currents at the $\sim 1 \times 10^{-28}$ e$\cdot$cm level.

\subsection{Field monitoring}

In order to control systematic uncertainties due to the motional magnetic field, it is essential to maintain stable, homogeneous electric field over the measurement cycle, as well as apply accurate electric-field reversal in order to reduce systematic effects quadratic in electric field. An accurate monitoring of electric field is necessary to ensure this has been achieved.

The Kerr effect in superfluid helium, already present in the experiment, provides a useful non-contact method of monitoring the electric field, especially given the harsh environment of the SNS nEDM experiment. The applied electric field causes the helium medium to become birefringent, which is then detected by laser polarimetry. The Kerr constant of superfluid helium has been measured to demonstrate the feasibility of the technique~\cite{sushkov}, and in order to minimize effect of spurious birefringence from optical windows, we have developed a double-pass cancellation scheme and demonstrated the scheme in a proof-of-concept setup~\cite{park}. We project sensitivity of $\delta E / E \approx 1\%$.

Non-linear magneto-optical rotation (NMOR) magnetometers can be used to monitor the magnetic field and ensure that the geometric phase effect from the magnetic field inhomogeneity is below the level specified above \cite{hovde}.

\section{Conclusion}
The EDM of elementary particles continue to provide important constraints on physics beyond the standard model.  Because they are
sensitive to CP-violating couplings, they are complementary to the direct searches at the LHC.  Using an array of techniques, some new, some old, a considerable number of experiments are either underway or being developed to improve the limits on the EDM of the electron, of nuclei, and of the neutron.  The neutron EDM experiment being developed for the Oak Ridge Spallation neutron source, using UCN production-in-place in superfluid $^4$He, has a goal of reaching into the $10^{-28}$ e$\cdot$cm regime with a roughly one year measurement time.

\end{document}